\documentstyle[axodraw,12pt]{article}          
\makeatletter  
\renewcommand{\theequation}{\thesection.\arabic{equation}}
\@addtoreset{equation}{section} \makeatother

\begin{document}

\renewcommand{\thefootnote}{\arabic{footnote}}
\newcommand{\hs}{\hspace*{0.5cm}}
\newcommand{\vs}{\vspace*{0.5cm}}
\newcommand{\be}{\begin{equation}}
\newcommand{\ee}{\end{equation}}
\newcommand{\bea}{\begin{eqnarray}}
\newcommand{\eea}{\end{eqnarray}}
\newcommand{\bary}{\begin{array}}
\newcommand{\eary}{\end{array}}
\newcommand{\bit}{\begin{itemize}}
\newcommand{\eit}{\end{itemize}}
\newcommand{\ben}{\begin{enumerate}}
\newcommand{\een}{\end{enumerate}}
\newcommand{\crn}{\nonumber \\}
\newcommand{\noi}{\noindent}
\newcommand{\al}{\alpha}
\newcommand{\la}{\lambda}
\newcommand{\bet}{\beta}
\newcommand{\ga}{\gamma}
\newcommand{\va}{\varphi}
\newcommand{\ci}[1]{\cite{#1}}
\newcommand{\re}[1]{(\ref{#1})}
\newcommand{\bi}[1]{\bibitem{#1}}
\newcommand{\lab}[1]{\label{#1}}
\newcommand{\ra}{\rightarrow}
\newcommand{\om}{\omega}
\newcommand{\pa}{\partial}
\newcommand{\fr}{\frac}
\newcommand{\sm}{\sigma}
\newcommand{\bc}{\begin{center}}
\newcommand{\ec}{\end{center}}
\newcommand{\nn}{\nonumber}
\newcommand{\Ga}{\Gamma}
\newcommand{\de}{\delta}
\newcommand{\De}{\Delta}
\newcommand{\ep}{\epsilon}
\newcommand{\varep}{\varepsilon}
\newcommand{\vthe}{\vartheta}
\newcommand{\ka}{\kappa}
\newcommand{\La}{\Lambda}
\newcommand{\vr}{\varrho}
\newcommand{\si}{\sigma}
\newcommand{\Si}{\Sigma}
\newcommand{\ta}{\tau}
\newcommand{\up}{\upsilon}
\newcommand{\Up}{\Upsilon}
\newcommand{\kh}{\chi}
\newcommand{\ze}{\zeta}
\newcommand{\ps}{\psi}
\newcommand{\Ps}{\Psi}
\newcommand{\ph}{\phi}
\newcommand{\vph}{\varphi}
\newcommand{\Ph}{\Phi}
\newcommand{\Om}{\Omega}

\def\lappeq{\mathrel{\rlap{\raise.5ex\hbox{$<$}}
{\lower.5ex\hbox{$\sim$}}}}

\bc {\Large\bf Fermion masses in the economical 3-3-1 model}

\vspace*{0.5cm}

P. V. Dong$^{a}$, D. T. Huong$^{a,b}$, T. T. Huong$^a$ and
H. N. Long$^{a,c}$ \\

\vspace*{0.5cm}

$^a$ {\it  Institute of  Physics, VAST, P. O. Box 429, Bo Ho,
Hanoi 10000, Vietnam}\\
$^b$ {\it Department of Physics and Astronomy, Aichi University of
Education,\\ Kariya 448-8542, Japan} $^c$ {\it The Abdus Salam
International Centre for Theoretical Physics,\\ 34014 Trieste,
Italy}

\ec

\begin{abstract}

We show that, in  frameworks of the economical 3-3-1 model, all
fermions get masses. At the tree level, one up-quark and two
down-quarks are massless, but the one-loop corrections give all
quarks the consistent masses. This conclusion is in contradiction to
the previous analysis in which, the third scalar triplet has been
introduced. This result is based on the key properties of the model:
First, there are three quite different scales of vacuum expectation
values: $\om \sim {\cal O}(1)\ \mathrm{TeV},\ v \approx 246\
\mathrm{GeV}$ and $ u \sim {\cal O}(1) \ \mathrm{GeV}$. Second,
there exist two types of Yukawa couplings with different strengths:
the lepton-number conserving couplings $h$'s and the lepton-number
violating ones $s$'s satisfying the condition in which the second
are much smaller than the first ones: $ s \ll h$.
 With the acceptable  set of parameters, numerical evaluation
 shows that in this model, masses of the exotic
quarks also have different scales, namely, the $U$ exotic  quark
($q_U = 2/3$) gains mass $m_U \approx 700 $ GeV, while the $D_\al$
exotic quarks ($q_{D_\al} = -1/3$) have masses in the TeV scale:
$m_{D_\al} \in 10 \div 80$ TeV.
\\

\noi PACS number(s): 12.15.Lk,  12.15.Ff,  11.30.Qc 

\end{abstract}

\noindent

\section{Introduction}

The recent experimental results of SuperKamiokande
Collaboration~\cite{superK},  KamLAND~\cite{kam} and SNO~\cite{sno}
confirm that neutrinos have tiny masses and  oscillate. This implies
that the standard model (SM) must
 be extended. Among the beyond-SM
extensions, the models based on the $\mathrm{SU}(3)_C\otimes
\mathrm{SU}(3)_L \otimes \mathrm{U}(1)_X$ (3-3-1) gauge group
\cite{ppf,flt} have some intriguing features: First, they can give
partial explanation of the generation number problem. Second,  the
third quark generation has to be different from the first two, so
this leads to the possible explanation of why top quark is
uncharacteristically heavy.

In one of the 3-3-1 models, the scalar sector is minimal with just
two Higgs triplets; hence it has been called the economical
model~\cite{haihiggs,ponce}. The general  Higgs sector of this model
is very simple (economical) and consists of three physical scalars
(two neutral and one charged) and eight Goldstone bosons - the
needed number for massive gauge ones~\cite{higgseconom}.

At the tree level, the mass matrix for the up-quarks has one
massless state and in the down-quark sector, there are two massless
ones. This calls radiative corrections. To solve this problem, the
authors in Ref. \cite{ponc1} have introduced the third Higgs
triplet. Therefore, it was though that the economical 3-3-1 model
with such only two Higgs triplets is not realistic.

In the present  work we will show that this is  a mistake! Without
the third one, at the one loop level, the fermions in this model,
with the given  set of parameters, gain a consistent mass spectrum.
A numerical evaluation leads us to conclusion that in this model,
there are two scales for masses of the exotic quarks.

The rest of this  paper is organized as follows: In Section
\ref{model}, we give a brief review of the economical 3-3-1 model.
Sec. \ref{fmass} is devoted for the fermion mass spectrum. In Sec.
\ref{oneloop} we present some details on the one-loop quark masses.
We summarize our result and make conclusions in the last section -
Sec. \ref{concl}.

\section{\label{model} A review of the economical 3-3-1 model}

The particle content in this model, which is anomaly free, is
given as follows:\bea \psi_{iL}&=&\left(%
\begin{array}{c}
  \nu_i \\
  e_i \\
  \nu^c_i \\
\end{array}%
\right)_L\sim \left(3,-\fr 1 3\right), \hs e_{iR}\sim (1,-1),\hs
i=1,2,3,\crn Q_{1L}&=&\left(%
\begin{array}{c}
  u_1 \\
  d_1 \\
  U \\
\end{array}%
\right)_L\sim \left(3,\fr 1 3\right),\hs Q_{\al L}=\left(%
\begin{array}{c}
  d_\al\\
  -u_\al\\
  D_\al\\
\end{array}%
\right)_L\sim (3^*,0),\hs \al=2,3,\crn u_{i R}&\sim&\left(1,\fr 2
3\right), \hs d_{i R} \sim \left(1,-\fr 1 3\right), \hs U_{R}\sim
\left(1,\fr 2 3\right),\hs D_{\al R} \sim \left(1,-\fr 1
3\right).\eea Here, the values in the parentheses denote quantum
numbers based on the $\left(\mathrm{SU}(3)_L,\mathrm{U}(1)_X\right)$
symmetry. In this case, the electric charge operator takes a form\be
Q=T_3-\fr{1}{\sqrt{3}}T_8+X,\label{eco}\ee where $T_a$
$(a=1,2,...,8)$ and $X$ stand for $\mathrm{SU}(3)_L$ and
$\mathrm{U}(1)_X$ charges, respectively.  Electric charges of the
exotic quarks $U$ and $D_\al$  are the same as of the usual quarks,
i.e., $q_{U}=2/3$ and $q_{D_\al}=-1/3$.

The $\mathrm{SU}(3)_L\otimes \mathrm{U}(1)_X$ gauge group is broken
spontaneously via two steps. In the first step, it is embedded in
that of the SM via a Higgs scalar triplet \be \chi=\left(%
\begin{array}{c}
  \chi^0_1 \\
  \chi^-_2 \\
  \chi^0_3 \\
\end{array}%
\right) \sim \left(3,-\fr 1 3\right)\ee acquired with a vacuum
expectation value (VEV) given by \be
\langle\chi\rangle=\fr{1}{\sqrt{2}}\left(%
\begin{array}{c}
  u \\
  0 \\
  \om \\
\end{array}%
\right).\label{vevc}\ee In the last step, to embed the gauge group
of the SM in $\mathrm{U}(1)_Q$, another Higgs scalar triplet \be
\phi=\left(%
\begin{array}{c}
  \phi^+_1 \\
  \phi^0_2 \\
  \phi^+_3 \\
\end{array}%
\right)\sim \left(3,\fr 2 3\right)\ee is needed with the VEV as
follows
\be \langle\phi\rangle=\fr{1}{\sqrt{2}}\left(%
\begin{array}{c}
  0 \\
  v \\
  0 \\
\end{array}%
\right).\label{vevp}\ee

The Yukawa interactions which induce masses for the fermions can be
written in the most general form as follows \be {\mathcal
L}_{\mathrm{Y}}={\mathcal L}_{\mathrm{LNC}} +{\mathcal
L}_{\mathrm{LNV}},\ee where the subscripts LNC and LNV, respectively
indicate to the lepton number conserving and lepton number violating
ones as shown below. Here, each part is defined by\bea {\mathcal
L}_{\mathrm{LNC}}&=&h^U\overline{Q}_{1L}\chi
U_{R}+h^D_{\al\beta}\overline{Q}_{\al L}\chi^* D_{\beta R}\crn
&&+h^e_{ij}\overline{\psi}_{iL}\phi
e_{jR}+h^\ep_{ij}\ep_{abc}(\overline{\psi}^c_{iL})_a(\psi_{jL})_b(\phi)_c
\crn && +h^d_{i}\overline{Q}_{1 L}\phi d_{i R}+h^u_{\al
i}\overline{Q}_{\al L}\phi^* u_{iR}+ H.c.,\label{y1}\\ {\mathcal
L}_{\mathrm{LNV}}&=&s^u_{i}\overline{Q}_{1L}\chi u_{iR}+s^d_{\al
i}\overline{Q}_{\al L}\chi^* d_{i R}\crn && +s^D_{
\al}\overline{Q}_{1L}\phi D_{\al R}+s^U_{\al }\overline{Q}_{\al
L}\phi^* U_{R}+ H.c.,\label{y2}\eea where $a$, $b$ and $c$ stand for
$\mathrm{SU}(3)_L$ indices.

The VEV $\om$ gives mass for the exotic quarks $U$ and $D_\al$, $u$
gives mass for $u_1, d_{\al}$ quarks, while $v$ gives mass for
$u_\al, d_{1}$ and {\it all} ordinary leptons. In the next sections
we provide more details on analysis of fermion masses. As mentioned,
the VEV $\om $ is responsible for the first step of symmetry
breaking, while the second step is due to $u$ and $v$. Therefore the
VEVs in this model have to be satisfied the constraint \be u, v \ll
\om . \label{vevcons} \ee

The Yukawa couplings of Eq.(\ref{y1}) possess an extra global
symmetry \cite{changlong,tujo} which is not broken by VEVs $ v,
\omega$ but by $u$. From these couplings, one can find the following
lepton symmetry $L$ as in Table \ref{lnumber} (only the fields with
nonzero $L$ are listed; all other fields have vanishing $L$). Here
$L$ is broken by $u$ which is behind $L(\chi^0_1)=2$, i.e., {\it $u$
is a kind of the lepton number violating parameter}.
\begin{table}[h]
 \caption{\label{lnumber} Nonzero lepton number $L$
 of the model particles.}
\bc
\begin{tabular}{c|c c c c c c c c c}
     \hline \\
        Field
&$\nu_{iL}$&$e_{iL,R}$&$\nu^c_{iL}$ & $\chi^0_1$&$\chi^-_2$ &
$\phi^+_3$ &
$U_{L,R}$ & $D_{2L,R}$& $D_{3L,R}$\\ \\
    \hline \\
        $L$ & $1$ & $1$ & $-1$ & $2$&$2$&$-2$&$-2$&$2$&$2$ \\ \\
        \hline
\end{tabular}
 \ec
\end{table} It is interesting that the exotic quarks also carry the lepton
number; therefore, this $L$ obviously does not commute with gauge
symmetry. One can construct a new conserved charge $\cal L$ through
$L$ by making the linear combination $L= xT_3 + yT_8 + {\cal L} I$.
Applying $L$ on a lepton triplet, the coefficients will be
determined \be L = \fr{4}{\sqrt{3}}T_8 + {\cal L} I \label{lepn}.\ee

Another useful conserved charge $\cal B$ which is exactly not broken
by $u$, $v$ and $\om$ is usual baryon number \be B ={\cal B} I.\ee
Both the $\mathcal{L}$ and $\mathcal{B}$ charges for the fermion and
Higgs multiplets are listed in Table~\ref{bcharge}.
\begin{table}[h]
\caption{\label{bcharge} ${\cal B}$ and ${\cal L}$ charges of the
model multiplets.} \bc
\begin{tabular}{c|cccccccccc}\hline\\
 Multiplet & $\chi$ & $\phi$ & $Q_{1L}$ & $Q_{\al L}$ &
$u_{iR}$&$d_{iR}$ &$U_R$ & $D_{\al R}$ & $\psi_{iL}$ & $e_{iR}$
\\ \\ \hline \\ $\cal B$-charge &$0$ & $ 0  $ &  $\fr 1 3  $ & $\fr 1 3
$& $\fr 1 3  $ &
 $\fr 1 3  $ &  $\fr 1 3  $&  $\fr 1 3  $&
 $0  $& $0   $\\ \\
\hline \\ $\cal L$-charge &$\fr 4 3$ & $-\fr 2 3  $ &
   $-\fr 2 3  $ & $\fr 2 3  $& 0 & 0 & $-2$& $2$&
 $\fr 1 3  $& $ 1   $\\ \\ \hline
\end{tabular}
 \ec
\end{table}

Let us note that the Yukawa couplings of (\ref{y2}) conserve
$\mathcal{B}$, while violate ${\mathcal L}$ with $\pm 2$ units which
implies that these interactions are much smaller than the first
ones, i.e., \be s_i^u, \ s_{\al i}^d,\ s_\al^D, \ s_\al^U \ll h^U,\
h_{\al \bet}^D,\ h_i^d,\ h_{\al i}^u.\label{dkhsyu}\ee In the
previous studies \cite{ponc1,violat}, the lepton number violating
terms of this kind have often been excluded, commonly by the
adoption of an appropriate discrete symmetry. There is no reason
within the 3-3-1 models why such terms should not be present. Let us
see next.

In this model, the most general Higgs potential has very simple form
\bea V(\chi,\phi) &=& \mu_1^2 \chi^\dag \chi + \mu_2^2 \phi^\dag
\phi + \la_1 ( \chi^\dag \chi)^2 + \la_2 ( \phi^\dag \phi)^2\crn & &
+ \la_3 ( \chi^\dag \chi)( \phi^\dag \phi) + \la_4 ( \chi^\dag
\phi)( \phi^\dag \chi). \label{poten} \eea Note that there is no
trilinear scalar coupling and this makes the Higgs potential much
simpler than the previous ones of the 3-3-1 models
\cite{changlong,tujo,ochoa2} and closer to that of the SM. The
analysis in Ref. \cite{higgseconom} shows that after symmetry
breaking, there are eight Goldstone bosons - needed number for
massive gauge ones, and three physical scalar fields. One of two
physical neutral scalars is the SM Higgs boson.

The non-zero values of $\chi$ and $\phi$ at the minimum value of
$V(\chi,\phi)$ can be obtained by\bea
\chi^+\chi&=&\fr{\lambda_3\mu^2_2
-2\lambda_2\mu^2_1}{4\lambda_1\lambda_2-\lambda^2_3}
\equiv\fr{u^2+\om^2}{2},\label{vev1}\\
\phi^+\phi&=&\fr{\lambda_3\mu^2_1
-2\lambda_1\mu^2_2}{4\lambda_1\lambda_2-\lambda^2_3}
\equiv\fr{v^2}{2}.\label{vev2}\eea Any other choice of $u,\om$ for
the vacuum value of $\chi$ satisfying (\ref{vev1}) gives the same
physics because it is related to (\ref{vevc}) by an
$\mbox{SU}(3)_L\otimes \mbox{U}(1)_X$ transformation. It is worth
noting that the assumed $u\neq 0$ is therefore given in a general
case.  This model of course leads to the formation of majorons
\cite{higgseconom}.  The constraint from
the $W$ decay width gives an upper limit on $t_\theta=u/\om$ which
leads to the fact that this LNV parameter can exceed, say
$t_\theta=0.08$ \cite{haihiggs}. Such a ratio of the scales for the
lepton number breaking is much larger than those in Refs.
\cite{tujo,plei}. Moreover, the Higgs potential (\ref{poten})
conserve both the mentioned global symmetries which implies that the
considering model explicitly differs from those in Refs.
\cite{changlong,tujo,ochoa2}.

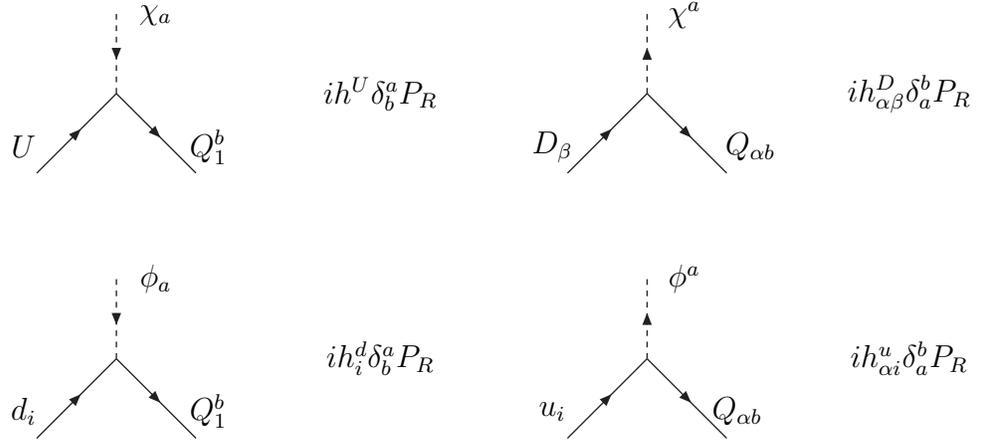
\begin{figure}[htbp]
 \bc
\begin{picture}(300,150)(0,-100)
\ArrowLine(0,10)(30,40)
 \ArrowLine(30,40)(60,10)
\DashArrowLine(30,70)(30,40){2}
 \Text(-5,20)[]{$U$}
 \Text(65,20)[]{$Q^b_{1}$}
  \Text(45,70)[]{$\chi_a$}
  \Text(130,40)[]{$ i h^U \de^a_b P_R $}
\ArrowLine(200,10)(230,40)
 \ArrowLine(230,40)(260,10)
 \DashArrowLine(230,40)(230,70){2}
 \Text(195,20)[]{$D_\beta$}
 \Text(270,20)[]{$Q_{\al b}$}
  \Text(245,70)[]{$\chi^a$}
  \Text(330,40)[]{$ i h^{D}_{\al \beta}\de^b_a P_R $}

\ArrowLine(0,-90)(30,-60)
 \ArrowLine(30,-60)(60,-90)
\DashArrowLine(30,-30)(30,-60){2}
 \Text(-5,-80)[]{$d_i$}
 \Text(65,-80)[]{$Q^b_1$}
  \Text(45,-30)[]{$\phi_a$}
  \Text(130,-60)[]{$ i  h^d_{i}\de^a_b P_R $}
\ArrowLine(200,-90)(230,-60)
 \ArrowLine(230,-60)(260,-90)
 \DashArrowLine(230,-60)(230,-30){2}
 \Text(195,-80)[]{$u_i$}
 \Text(265,-80)[]{$Q_{\al b}$}
  \Text(245,-30)[]{$\phi^a$}
  \Text(330,-60)[]{$ i  h^u_{\al i}\de^b_a P_R $}
\end{picture}
\ec
 \caption[]{\label{figh1} The relevant lepton-number conserving
 couplings to quarks}
\end{figure}

\begin{figure}[htbp]
 \bc
\begin{picture}(300,150)(0,-100)
\ArrowLine(0,10)(30,40)
 \ArrowLine(30,40)(60,10)
\DashArrowLine(30,70)(30,40){2}
 \Text(-5,20)[]{$u_i$}
 \Text(65,20)[]{$Q^b_{1}$}
  \Text(45,70)[]{$\chi_a$}
  \Text(130,40)[]{$ i s^u_{i} \de^a_b P_R $}
\ArrowLine(200,10)(230,40)
 \ArrowLine(230,40)(260,10)
 \DashArrowLine(230,40)(230,70){2}
 \Text(195,20)[]{$d_i$}
 \Text(270,20)[]{$Q_{\al b}$}
  \Text(245,70)[]{$\chi^a$}
  \Text(330,40)[]{$ i s^{d}_{\al i}\de^b_a P_R $}

\ArrowLine(0,-90)(30,-60)
 \ArrowLine(30,-60)(60,-90)
\DashArrowLine(30,-30)(30,-60){2}
 \Text(-5,-80)[]{$D_\al$}
 \Text(65,-80)[]{$Q^b_1$}
  \Text(45,-30)[]{$\phi_a$}
  \Text(130,-60)[]{$ i  s^D_{\al}\de^a_b P_R $}
\ArrowLine(200,-90)(230,-60)
 \ArrowLine(230,-60)(260,-90)
 \DashArrowLine(230,-60)(230,-30){2}
 \Text(195,-80)[]{$U$}
 \Text(269,-80)[]{$Q_{\al b}$}
  \Text(245,-30)[]{$\phi^a$}
  \Text(330,-60)[]{$ i s^U_{\al }\de^b_a P_R $}
\end{picture}
\ec
 \caption[]{\label{figh2} The lepton-number violating
 couplings}
\end{figure}
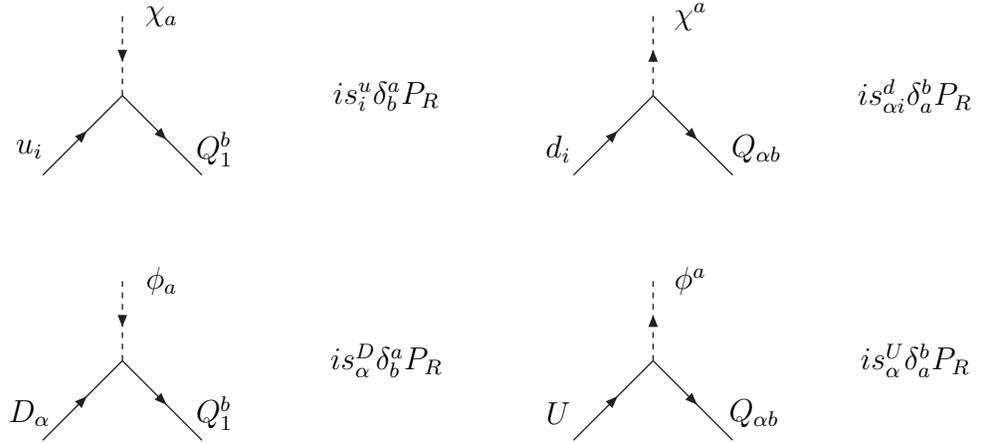

For the sake of convenience in further reading, we present the
relevant Yukawa couplings in terms of Feynman diagrams in the
figures (\ref{figh1}) and (\ref{figh2}), where the
Hermitian-conjugated couplings are not displayed. \vs

\vs

\vs

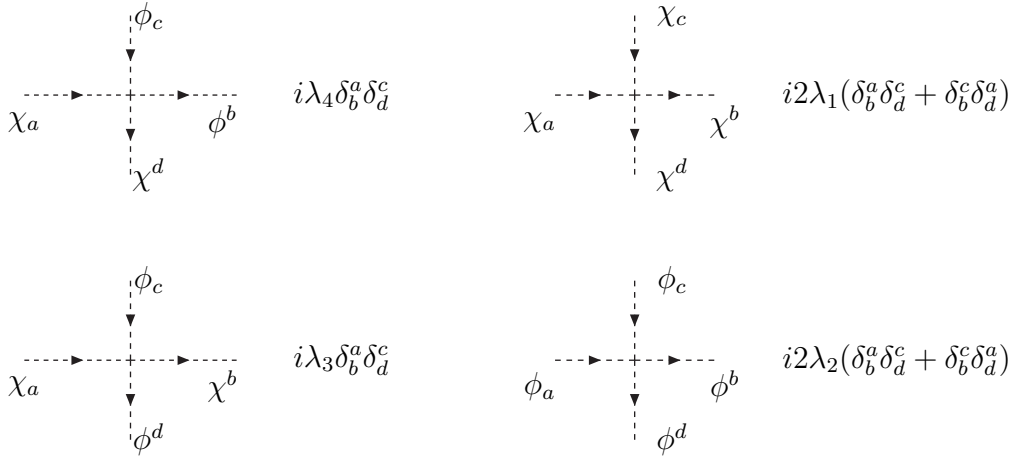
\begin{figure}[t]
\bc
\begin{picture}(300,180)(10,-100)

\DashArrowLine(0,40)(40,40){2}
 \DashArrowLine(40,40)(80,40){2}
 \DashArrowLine(40,70)(40,40){2}
 \DashArrowLine(40,40)(40,10){2}

 \Text(0,30)[]{$\chi_a$}
 \Text(75,30)[]{$\phi^b$}
  \Text(120,40)[]{$i\la_4 \de_b^a \de_d^c $}
 \Text(47,70)[]{$\phi_c$}
\Text(47,10)[]{$\chi^d$}
\DashArrowLine(230,40)(230,10){2} \DashArrowLine(200,40)(230,40){2}
 \DashArrowLine(230,40)(260,40){2}
 \DashArrowLine(230,70)(230,40){2}
 \Text(195,30)[]{$\chi_a$}
 \Text(265,30)[]{$\chi^b$}
  \Text(245,70)[]{$\chi_c$}
    \Text(245,10)[]{$\chi^d$}
  \Text(330,40)[]{$ i 2\la_{1}(\de_b^a \de_d^c
  + \de_b^c \de_d^a) $}

\DashArrowLine(0,-60)(40,-60){2}
 \DashArrowLine(40,-60)(80,-60){2}
 \DashArrowLine(40,-30)(40,-60){2}
 \DashArrowLine(40,-60)(40,-90){2}

 \Text(0,-70)[]{$\chi_a$}
 \Text(75,-70)[]{$\chi^b$}
  \Text(120,-60)[]{$i\la_3 \de_b^a \de_d^c $}
 \Text(47,-30)[]{$\phi_c$}
\Text(47,-90)[]{$\phi^d$}

\DashArrowLine(230,-60)(230,-90){2}
\DashArrowLine(200,-60)(230,-60){2}
 \DashArrowLine(230,-60)(260,-60){2}
 \DashArrowLine(230,-30)(230,-60){2}
 \Text(195,-70)[]{$\phi_a$}
 \Text(265,-70)[]{$\phi^b$}
  \Text(245,-30)[]{$\phi_c$}
    \Text(245,-90)[]{$\phi^d$}
  \Text(330,-60)[]{$ i 2\la_{2}(\de_b^a \de_d^c
  + \de_b^c \de_d^a) $}
\end{picture}
\ec
 \caption[]{The  Higgs boson self-couplings   }
\label{figh3}
\end{figure}

The Higgs boson self-couplings
are depicted in the figure (\ref{figh3}).\vs

\vs

Let us note that in Ref. \cite{ponc1}, the authors have considered
the fermion mass spectrum under the $Z_2$ discrete symmetry which
discards the LNV interactions. Here the couplings of the figure
(\ref{figh2}) in such case are forbidden.  Then it can be checked
that some quarks remain massless up to two-loop level. To solve the
mass problem of the quarks, the authors in Ref. \cite{ponc1} have
shown that one third scalar triplet has to be added to the resulting
model.

In the following we show that it is not necessary. The $Z_2$ is not
introduced and thus the third one is not required. The LNV Yukawa
couplings are vital for the economical 3-3-1 model.

\section{\label{fmass} Fermion masses}

The VEV $\om$ breaks $\mathrm{SU}(3)_L \otimes \mathrm{U}(1)_X$ down
to $\mathrm{SU}(2)_L \otimes \mathrm{U}(1)_Y$ and gives masses of
the exotic quarks as well as non-SM gauge bosons $X,Y$ and $Z'$. The
VEV $u$ gives mass for $u_1, d_{\al}$ quarks, while $v$ gives mass
for $u_\al, d_{1}$ and all ordinary leptons. The SM gauge bosons
gain mass both from $u$ and $v$. The fermions gain mass terms via
Yukawa interactions given in (\ref{y1}) and (\ref{y2}).

\subsection{Lepton masses}
The charged leptons $(l= e_i=e, \mu, \tau)$ gain masses via the
following couplings \bea {\cal L}^e_Y= h^e_{ij}\overline{\psi}_{iL}
\phi e_{jR}+ \mathrm{H.c.}\eea
The mass matrix is therefore followed by \bea
M_{l}= -\fr{v}{\sqrt{2}}\left(%
\begin{array}{ccc}
  h^e_{11} & h^e_{12} & h^e_{13} \\
  h^e_{21} & h^e_{22} & h^e_{23} \\
  h^e_{31} & h^e_{32} & h^e_{33} \\
\end{array}%
\right),\eea which of course gives consistent masses for the charged
leptons \cite{ponc1}.

Interesting physics is of the neutrino sector. At the tree level,
the neutrinos gain Dirac masses via\bea M_\nu=-\sqrt{2}v\left(%
\begin{array}{ccc}
  0 & h^\ep_{12} & h^\ep_{13} \\
  -h^\ep_{12} & 0 & h^\ep_{23} \\
  -h^\ep_{13} & -h^\ep_{23} & 0 \\
\end{array}%
\right),\eea which gives the mass pattern $0,-m_\nu,m_\nu$, here
$m_\nu \equiv v\sqrt{2(h^{\ep 2}_{12}+h^{\ep 2}_{13}+h^{\ep
2}_{23})}$. This means that one neutrino is massless and two are
degenerate in mass. This pattern is clearly not realistic under the
current data; however, it is severely changed by the quantum effect.
We will return on this topics in the future publication.

\subsection{Quark  masses} \label{treelevel}

The Yukawa couplings in (\ref{y1}) and (\ref{y2}) give the mass
Lagrangian for the up-quarks (quark sector with electric charge
$q_{\mathrm{up}}=2/3$) \bea
\mathcal{L}_{\mathrm{up}}^{\mathrm{mass}} &=&
\fr{h^U}{\sqrt{2}}\left( \overline{u}_{1L} u + \overline{U}_L \om
\right) U_R + \fr{s^u_{i}}{\sqrt{2}}\left( \overline{u}_{1L} u +
\overline{U}_L \om \right) u_{iR} \crn && - \fr{v }{\sqrt{2}}
\overline{u}_{\al L}\left(  h^u_{\al i} u_{iR} + s^U_{\al } U_R
\right) + H. c. \label{upqmasst} \eea
Consequently, we obtain the mass matrix for the up-quarks  $(u_1,
u_2, u_3, U)$ as follows \be
M_{\mathrm{up}} = \fr{1}{\sqrt{2}}\left(%
\begin{array}{cccc}
  -s^u_{1} u & -s^u_{2}u  & -s^u_{3}u & -h^U u \\
 h^u_{2 1} v  & h^u_{2 2} v & h^u_{2 3} v & s^U_{2} v \\
 h^u_{3 1} v  & h^u_{3 2} v & h^u_{3 3} v & s^U_{3} v  \\
  -s^u_{1} \om  & -s^u_{2} \om & -s^u_{3} \om & -h^U \om \\
\end{array}%
\right) \label{upqmasstu} \ee Because the first and the last rows of
the matrix (\ref{upqmasstu}) are proportional, the tree level
up-quark spectrum contains a massless one!

Similarly,  for the down-quarks ($q_{\mathrm{down}}= -1/3$), we get
the following mass Lagrangian \bea
\mathcal{L}_{\mathrm{down}}^{\mathrm{mass}} &=& \fr{h^D_{\al
\bet}}{\sqrt{2}}\left( \overline{d}_{\al L} u + \overline{D}_{\al L}
\om \right) D_{\bet R} + \fr{s^d_{\al i}}{\sqrt{2}}\left(
\overline{d}_{\al L} u + \overline{D}_{\al L} \om \right) d_{iR}
\crn && + \fr{v }{\sqrt{2}} \overline{d}_{1 L}\left(  h^d_{ i}
d_{iR} + s^D_{\al} D_{\al R} \right) + H. c. \label{upqmassdow} \eea
Hence we get mass matrix for the down-quarks $(d_1, d_2, d_3, D_2,
D_3)$ \be M_{\mathrm{down}} = - \fr{1}{\sqrt{2}}
\left(%
\begin{array}{ccccc}
  h^d_{1} v  &  h^d_{2} v  &  h^d_{3} v  &  s^D_{2} v  & s^D_{3} v \\
  s^d_{21} u &  s^d_{22} u &  s^d_{23} u &  h^D_{22} u &  h^D_{23} u \\
   s^d_{31} u &  s^d_{32} u &  s^d_{33} u  & h^D_{32} u  & h^D_{33} u  \\
   s^d_{21} \om & s^d_{22} \om & s^d_{23} \om & h^D_{22} \om & h^D_{23} \om  \\
  s^d_{31} \om & s^d_{32} \om & s^d_{33} \om & h^D_{32} \om & h^D_{33}\om \\
\end{array}%
\right)
 \label{upqmasstdow2} \ee
We see that the second and fourth rows of matrix in
(\ref{upqmasstdow2})   are proportional, while the third and the
last are the same. Hence, in this case there are two massless
eigenstates.

The masslessness of the tree level quarks in both the sectors calls
radiative corrections (the so-called mass problem of quarks). These
corrections start at the one-loop level. The diagrams in the figure
(\ref{figh4}) contribute the up-quark spectrum while the figure
(\ref{figh5}) gives the down-quarks. Let us note the reader that the
quarks also get some one-loop contributions in the case of the $Z_2$
symmetry enforcing \cite{ponc1}. The careful study of this radiative
mechanism shows that the one-loop quark spectrum is consistent.

\section{Typical examples of the one-loop corrections}
\label{oneloop}

To provide the quarks masses, in the following we can suppose that
the Yukawa couplings are flavor diagonal. Then the $u_2$ and $u_3$
states are mass eigenstates corresponding to the mass eigenvalues:
\be m_2= h^u_{22}\fr{v}{\sqrt{2}},\hs
m_{3}=h^u_{33}\fr{v}{\sqrt{2}}. \label{hh1} \ee The $u_1$ state
mixes with the exotic $U$ in terms of one sub-matrix of the mass
matrix (\ref{upqmasst})
\be M_{uU}=-\fr{1}{\sqrt{2}}\left(%
\begin{array}{cc}
  s^u_1 u & h^U u \\
  s^u_1 \om & h^U \om \\
\end{array}%
\right).\label{m11}\ee This matrix contains one massless quark $\sim
u_1$, $m_1=0$, and the remaining exotic quark $\sim U$ with the mass
of the scale $\om$.

Similarly, for the down-quarks, the $d_1$ state is a mass eigenstate
corresponding to the eigenvalue: \be m'_1 =
-h^d_{1}\fr{v}{\sqrt{2}}.\ee The pairs $(d_2,D_2)$ and $(d_3,D_3)$
are decouple, while the quarks of each pair mix via the mass
sub-matrices, respectively,\bea M_{d_2D_2}&=&
-\fr{1}{\sqrt{2}}\left(%
\begin{array}{cc}
  s^d_{22} u & h^D_{22} u \\
  s^d _{22}\om & h^D_{22} \om \\
\end{array}%
\right),\label{m22}\\ M_{d_3D_3}&=&-\fr{1}{\sqrt{2}}\left(%
\begin{array}{cc}
  s^d_{33} u & h^D_{33} u \\
  s^d _{33}\om & h^D_{33} \om \\
\end{array}%
\right).\label{m33}\eea These matrices contain the massless quarks
$\sim d_2$ and $d_3$ corresponding to $m'_2=0$ and $m'_3=0$, and two
exotic quarks $\sim D_2$ and $D_3$ with the masses of the scale
$\om$.

 With the help of the constraint (\ref{vevcons}), we identify
$m_1$, $m_2$ and $m_3$ respective to those of the $u_1=u$, $u_2 = c$
and $u_3 = t$ quarks. The down quarks $d_1$, $d_2$ and $d_3$ are
therefore corresponding to $d$, $s$ and $b$ quarks. Unlike the usual
3-3-1 model with right-handed neutrinos, where the third family of
quarks should be discriminating \cite{longvan}, in the model under
consideration the {\it first} family has to be different from the
two others.

The mass matrices (\ref{m11}), (\ref{m22}) and (\ref{m33}) remain
the tree level properties for the quark spectra - one massless in
the up-quark sector and two in the down-quarks. From these matrices,
it is easily to verify that the conditions in (\ref{vevcons}) and
(\ref{dkhsyu}) are satisfied. First, we consider radiative
corrections to the up-quark masses.

\subsection{Up quarks} \label{treelevel}

In the previous studies \cite{ponc1,violat}, the LNV interactions
have often been excluded, commonly by the adoption of an appropriate
discrete symmetry. Let us remind that there is no reason within the
3-3-1 model to ignore such interactions. The experimental limits on
processes which do not conserve total lepton numbers, such as
neutrinoless double beta decay \cite{0bedecay}, require them to be
small.

\begin{figure}[htbp]
\begin{center}
\begin{picture}(650,120)(40,-150)
\ArrowLine(210,-120)(240,-120) \ArrowLine(330,-120)(360,-120)
\ArrowLine(240,-120)(285,-120) \ArrowLine(285,-120)(330,-120)
\DashArrowArcn(285,-120)(45,90,0){2}
\DashArrowArc(285,-120)(45,90,180){2} \Text(350,-130)[]{$u_{1L}$}
\Text(210,-130)[]{$ U_{R} $} \Text(260,-130)[]{$ U_{L}$}
\Text(305,-130)[]{$ U_{R}$} \Text(240,-130)[]{$h^U$}
\Text(285,-120)[]{$\times$} \Text(285,-110)[]{$\om$}
\Text(330,-130)[]{$h^U $}
\Text(333,-95)[]{$ \chi^{0}_1$} 
\Text(244,-90)[]{$ \chi^0_3$} \DashArrowLine(250,-40)(285,-75){2}
\DashArrowLine(320,-40)(285,-75){2} \Text(320,-40)[]{$\times$}
\Text(250,-40)[]{$\times$} \Text(250,-60)[]{$\chi^{0}_1$}
\Text(320,-60)[]{$\chi^{0}_3$}
\Text(285,-85)[]{$  \la_{1}$}
\end{picture}
\end{center}
\caption[]{\label{figh6} One-loop contribution under $Z_2$ to the
up-quark mass matrix (\ref{m110})}
\end{figure}
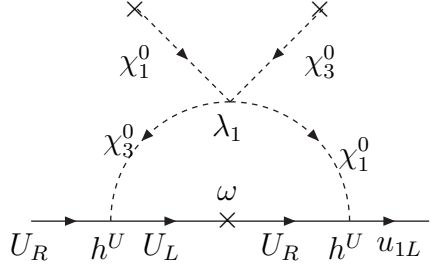

If the Yukawa Lagrangian is restricted to
$\mathcal{L}_{\mathrm{LNC}}$ \cite{ponc1}, then the mass matrix
(\ref{m11}) becomes
\be M_{uU}=-\fr{1}{\sqrt{2}}\left(%
\begin{array}{cc}
  0 & h^U u \\
  0 & h^U \om \\
\end{array}%
\right).\label{m110}\ee In this case, only the element
$(M_{uU})_{12}$ gets an one-loop correction defined by the figure
(\ref{figh6}). Other elements remain unchanged under this one-loop
effect.\vs

The Feynman rules gives us \bea -i (M_{uU})_{12}P_R &=&\int \fr{d^4
p}{(2\pi)^4}(ih^U P_R)\fr{i(p\!\!\!/+M_{U})}{p^2-M^2_{U}}(- i M_U
P_L) \fr{i(p\!\!\!/+M_{U})}{p^2-M^2_{U}}(ih^U P_R)\crn &
&\times\fr{-1}{(p^2-M^2_{\chi_1})(p^2-M^2_{\chi_3})}(i 4 \la_1)\fr{
u\om}{2}.\nn\eea  Thus, we get \bea (M_{uU})_{12} &=&-2i u\om \la_1
M_{U} (h^U)^2 \int \fr{d^4
p}{(2\pi)^4}\fr{p^2}{(p^2-M^2_U)^2(p^2-M^2_{\chi_3})(p^2-M^2_{\chi_1})}
\crn &\equiv&-2i u\om \la_1 M_{U} (h^U)^2
I(M^2_U,M^2_{\chi_3},M^2_{\chi_1}).\eea The integral $I(a,b,c)$ with
$a,b\gg c$ is given in  the  \ref{inter}.  Following  Ref.
\cite{higgseconom},  we conclude that in an effective approximation,
$M^2_U,\ M^2_{\chi_3} \gg M^2_{\chi_1}$. Hence we have \bea
(M_{uU})_{12} &\simeq&-\fr{\la_1 t_\theta  M^3_{U}
}{4\pi^2}\left[\fr{M^2_U-M^2_{\chi_3}+M^2_{\chi_3}\ln\fr{
M^2_{\chi_3}}{M^2_U}}{(M^2_U-M^2_{\chi_3})^2}\right]\sim u, \crn
&\equiv& -\fr{1}{\sqrt{2}}R(M_U). \eea The resulting mass matrix is
given by
\be M_{uU}=-\fr{1}{\sqrt{2}}\left(%
\begin{array}{cc}
  0 & h^U u+R \\
  0 & h^U \om \\
\end{array}%
\right).\label{m111}\ee We see that one quark remains massless as
the case of the tree level spectrum. This result keeps up to
two-loop level, and can be applied to the down-quark sector as well
as in the cases of non-diagonal Yukawa couplings. Therefore, under
the $Z_2$, it is not able to provide consistent masses for the
quarks.

If the full Yukawa Lagrangian is used, the LNV couplings must be
enough small in comparison with the usual couplings   [see
(\ref{dkhsyu})]. Combining (\ref{vevcons}) and (\ref{dkhsyu}) we
have \be h^U\om \gg h^U u,\ s^u_1 \om \gg s^u_1 u.\ee In this case,
the element  $(M_{uU})_{11}$ of (\ref{m11}) gets the radiative
correction  depicted in Fig.(\ref{figh7}). \vs
\begin{figure}[htbp]
\begin{center}
\begin{picture}(600,120)(30,-20)
\ArrowLine(210,20)(240,20) \ArrowLine(330,20)(360,20)
\ArrowLine(240,20)(285,20) \ArrowLine(285,20)(330,20)
\DashArrowArcn(285,20)(45,90,0){2}
\DashArrowArc(285,20)(45,90,180){2} \Text(350,10)[]{$u_{1L}$}
\Text(210,10)[]{$ u_{1R} $} \Text(260,10)[]{$ U_{L}$}
\Text(305,10)[]{$ U_{R}$} \Text(240,10)[]{$s^u_{1}$}
\Text(285,20)[]{$\times$} \Text(285,30)[]{$\om$}
 \Text(330,10)[]{$h^U $}
\Text(333,45)[]{$ \chi^{0}_1$} 
\Text(244,50)[]{$ \chi^0_3$} \DashArrowLine(250,100)(285,65){2}
\DashArrowLine(320,100)(285,65){2} \Text(320,100)[]{$\times$}
\Text(250,100)[]{$\times$} \Text(250,80)[]{$\chi^{0}_1$}
\Text(320,80)[]{$\chi^{0}_3$}
\Text(285,55)[]{$
\la_{1}$}
\end{picture}
\end{center}
\caption[]{\label{figh7} One-loop contribution to the up-quark mass
matrix (\ref{m11})}
\end{figure} The resulting mass matrix is obtained by
\be M_{uU}=-\fr{1}{\sqrt{2}}\left(%
\begin{array}{cc}
  s^u_1( u +\fr{R}{h^U}) & h^U u \\
  s^u_1 \om & h^U \om \\
\end{array}%
\right).\label{m1111}\ee In contradiction with the first case, the
mass of $u$ quark is now non-zero and given by\be m_u \simeq
\fr{s^u_1}{\sqrt{2}h^U} R \label{mu2}.\ee

Let us note that the matrix (\ref{m1111}) gives an eigenvalue in the
scale of $\fr{1}{\sqrt{2}}h^U \om$ which can be identified with that
of the exotic quark $U$. In effective approximation
\cite{higgseconom}, the mass for the Higgs $\chi_3$ is defined by
$M^2_{\chi_3}\simeq 2\la_1 \om^2$. Hereafter, for the parameters, we
use the following values $\la_1=2.0$, $t_\theta=0.08$ as mentioned,
and $\om =10\ \mathrm{TeV}$. The mass value for the $u$ quark is as
function of $s^u_1$ and $h^U$. Some values of the pair $(s^u_1,h^U)$
which give consistent masses for the $u$ quark is listed in Table
\ref{uvalues}.
\begin{table}[h]
\caption{\label{uvalues} Mass for the $u$ quark as function of
$(s^u_1,h^U)$.}\bc
\begin{tabular}{c|ccccc}
  \hline \\
  $h^U$ & 2 & 1.5 & 1 & 0.5 & 0.1\\ \\
  $s^u_1$& 0.0002 & 0.0003 & 0.0004 & 0.001 & 0.01 \\ \\
    \hline \\
  $m_u$ [MeV]& 2.207 & 2.565 & 2.246 & 2.375 & 2.025 \\ \\
  \hline
\end{tabular}
\ec
\end{table}

Note that the mass values in the Table \ref{uvalues} for the $u$
quark are in good consistence with the data given in
Ref.~\cite{pdg}: $m_u \in 1.5\ \div \ 4 \ \mathrm{MeV}$.

\subsection{Down quarks}

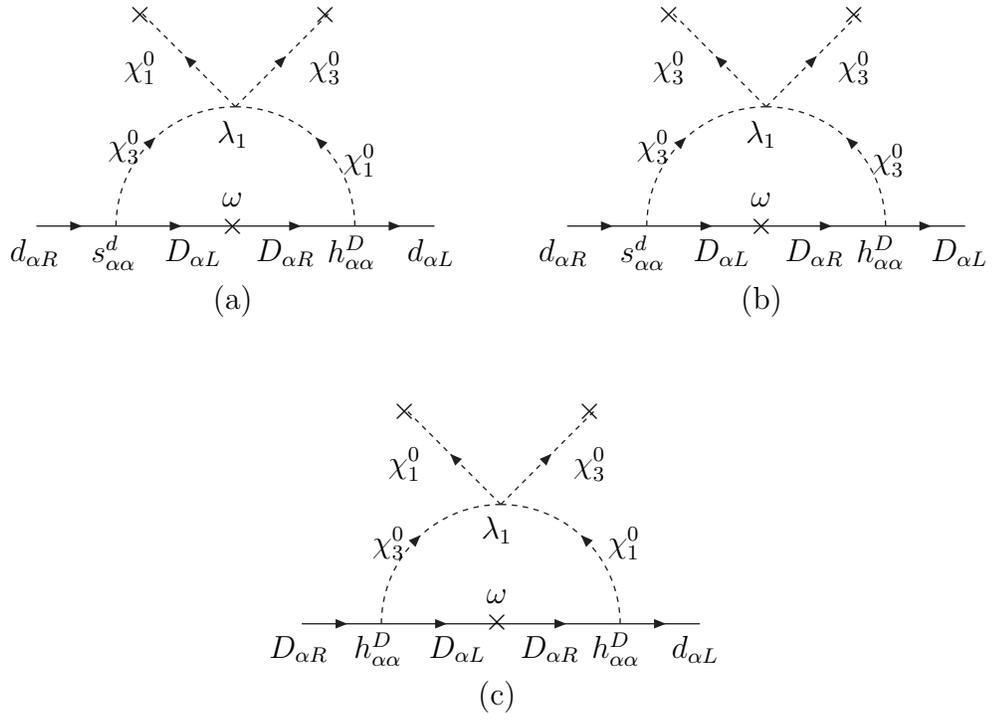
\begin{figure}[htbp]
\begin{center}
\begin{picture}(600,300)(150,-200)
\ArrowLine(210,20)(240,20) \ArrowLine(330,20)(360,20)
\ArrowLine(240,20)(285,20) \ArrowLine(285,20)(330,20)
\DashArrowArc(285,20)(45,0,90){2}
\DashArrowArcn(285,20)(45,180,90){2} \Text(360,10)[]{$d_{\al L}$}
\Text(210,10)[]{$ d_{\al R} $} \Text(270,10)[]{$ D_{\al L}$}
\Text(305,10)[]{$ D_{\al R}$} \Text(240,10)[]{$s^d_{\al \al}$}
\Text(285,20)[]{$\times$} \Text(285,30)[]{$\om$}
 \Text(330,10)[]{$h^D_{\al \al} $}
\Text(333,45)[]{$ \chi^{0}_1$} 
\Text(244,50)[]{$ \chi^0_3$} \DashArrowLine(285,65)(250,100){2}
\DashArrowLine(285,65)(320,100){2} \Text(320,100)[]{$\times$}
\Text(250,100)[]{$\times$} \Text(250,80)[]{$\chi^{0}_1$}
\Text(320,80)[]{$\chi^{0}_3$}\Text(285,-8)[]{(a)} \Text(285,55)[]{$
\la_{1}$}
\ArrowLine(410,20)(440,20) \ArrowLine(530,20)(560,20)
\ArrowLine(440,20)(485,20) \ArrowLine(485,20)(530,20)
\DashArrowArc(485,20)(45,0,90){2}
\DashArrowArcn(485,20)(45,180,90){2} \Text(560,10)[]{$D_{\al L}$}
\Text(410,10)[]{$ d_{\al R} $} \Text(470,10)[]{$ D_{\al L}$}
\Text(505,10)[]{$ D_{\al R}$} \Text(440,10)[]{$s^d_{\al \al}$}
\Text(485,20)[]{$\times$} \Text(485,30)[]{$\om$}
 \Text(530,10)[]{$h^D_{\al \al} $}
\Text(533,45)[]{$ \chi^0_3$} 
\Text(444,50)[]{$ \chi^0_3$} \DashArrowLine(485,65)(450,100){2}
\DashArrowLine(485,65)(520,100){2} \Text(520,100)[]{$\times$}
\Text(450,100)[]{$\times$} \Text(450,80)[]{$\chi^{0}_3$}
\Text(520,80)[]{$\chi^{0}_3$}\Text(485,-8)[]{(b)} \Text(485,55)[]{$
\la_{1}$}
\ArrowLine(310,-130)(340,-130) \ArrowLine(430,-130)(460,-130)
\ArrowLine(340,-130)(385,-130) \ArrowLine(385,-130)(430,-130)
\DashArrowArc(385,-130)(45,0,90){2}
\DashArrowArcn(385,-130)(45,180,90){2} \Text(460,-140)[]{$d_{\al
L}$} \Text(310,-140)[]{$ D_{\al R} $} \Text(370,-140)[]{$ D_{\al
L}$} \Text(405,-140)[]{$ D_{\al R}$} \Text(340,-140)[]{$h^D_{\al
\al}$} \Text(385,-130)[]{$\times$} \Text(385,-120)[]{$\om$}
 \Text(430,-140)[]{$h^D_{\al \al} $}
\Text(433,-100)[]{$ \chi^0_1$} 
\Text(344,-100)[]{$ \chi^0_3$} \DashArrowLine(385,-85)(350,-50){2}
\DashArrowLine(385,-85)(420,-50){2} \Text(420,-50)[]{$\times$}
\Text(350,-50)[]{$\times$} \Text(350,-70)[]{$\chi^{0}_1$}
\Text(420,-70)[]{$\chi^{0}_3$}\Text(385,-158)[]{(c)}
\Text(385,-95)[]{$  \la_{1}$}

\end{picture}
\end{center}
\caption[]{\label{figh8} One-loop contributions to the down-quark
mass matrix (\ref{m22}) or (\ref{m33}).}
\end{figure}

For the down quarks, the constraint, \be h^D_{\al\al} \om \gg
h^D_{\al\al} u,\ s^d_{\al\al} \om \gg s^d_{\al \al} u,\ee should be
applied. In this case, three elements $(M_{d_\al D_\al})_{11}$,
$(M_{d_\al D_\al})_{12}$ and $(M_{d_\al D_\al})_{21}$ will get
radiative corrections. The relevant diagrams are depicted in figure
(\ref{figh8}). \vs

It is worth noting that diagram \ref{figh8}(c) exists even in the
case of the $Z_2$ symmetry. The contributions are given by \bea
(M_{d_\al D_\al})_{11}&=&-\fr{s^d_{\al \al}}{\sqrt{2}h^D_{\al
\al}}R(M_{D_\al}),\\ (M_{d_\al D_\al})_{21}&=&-4i \la_1 \fr{s^d_{\al
\al}}{h^D_{\al \al}}M^3_{D_{ \al}}
I(M^2_{D_{\al}},M^2_{\chi_3},M^2_{\chi_3})\crn &=&-\fr{\la_1
s^d_{\al\al}M^3_{D_{ \al}}}{4 \pi^2 h^D_{\al
\al}}\left[\fr{M^2_{D_{\al}}+M^2_{\chi_3}}{(M^2_{D_{\al}}
-M^2_{\chi_3})^2}-\fr{2M^2_{D_{\al}}M^2_{\chi_3}}
{(M^2_{D_{\al}}-M^2_{\chi_3})^3}\ln\fr{M^2_{D_{\al}}}
{M^2_{\chi_3}}\right]\crn &\equiv&-\fr{1}{\sqrt{2}}R'(M_{D_{\al}}),
\\ (M_{d_\al D_\al})_{12}&=&-\fr{1}{\sqrt{2}}R(M_{D_\al}).\eea
We see that two last terms are much larger than the first one. This
is responsible for the masses of the quarks $d_2$ and $d_3$. At the
one-loop level, the mass matrix for the down-quarks is given by
 \be M_{d_\al D_\al}=
-\fr{1}{\sqrt{2}}\left(%
\begin{array}{cc}
  s^d_{\al \al}( u+\fr{R}{h^D_{\al\al}}) & h^D_{\al\al} u+R \\
  s^d _{\al\al}\om + R' & h^D_{\al\al} \om \\
\end{array}%
\right).\label{domass}\ee

We remind the reader that a matrix (see also \cite{chli})\be \left(
  \begin{array}{cc}
    a & c \\
    b & D \\
  \end{array}
\right) \label{ptbac2}\ee with $D\gg b, c \gg a$ has two eigenvalues
\bea x_1 &\simeq&
\left[a^2-\fr{2bca}{D}+\fr{b^2c^2-(b^2+c^2)a^2}{D^2}\right]^{1/2},\crn
x_2 & \simeq & D. \label{ptbac22}\eea Therefore
 the mass matrix in (\ref{domass}) gives an eigenvalue in the scale of
$D \equiv \fr{1}{\sqrt{2}} h^D_{\al\al} \om $ which is of the exotic
quark $D'_\al$. Here we  have another eigenvalue for the mass of
$d'_\al$ \be m_{d'_\al}
=\fr{h^D_{\al\al}u+R}{\sqrt{2}h^D_{\al\al}\om}\left\{R'^2
-\fr{(s^d_{\al\al})^2}{(h^D_{\al\al})^2}\left[(s^d_{\al
\al}\om+R')^2+(h^D_{\al \al}u+R)^2\right]\right\}^{1/2}
\label{ptbac23}.\ee

Let us remember that $M^2_{\chi_3}\simeq 2\la_1 \om^2$, and the
parameters $\la_1=2.0$, $t_\theta=0.08$ and $\om =10\ \mathrm{TeV}$
as given above are used in this case. The $m_{d_\al}$ is function of
$s^d_{\al\al}$ and $h^D_{\al\al}$. We take the value
$h^D_{\al\al}=2.0$ for both the sectors, $\al=2$ and $\al=3$. If
$s^d_{22}=0.015$ we get then the mass of the so-called $s$ quark \be
m_s = 99.3\ \mathrm{MeV}.\label{ms1}\ee For the down quark of the
third family, we put $s^d_{33}=0.7$. Then, the mass of the $b$ quark
is obtained by \be m_b =4.4\ \mathrm{GeV}.\label{mb1}\ee

We emphasize again  that Eqs. (\ref{ms1}) and (\ref{mb1}) are in
good consistence with the data given in Ref.~\cite{pdg}: $m_s \sim
95\ \pm 25 \ \mathrm{MeV}$ and  $m_b \sim 4.70 \pm 0.07 \
\mathrm{GeV}$.

\section{Summary and Conclusions}
\label{concl} In this paper,  we have presented the answer to one of
the most crucial questions: whether within the minimal scalar Higgs
content, all fermions including quarks can gain the consistent
masses.

In Ref.~\cite{haihiggs} we have shown that, in the considered model,
there are three quite different scales of vacuum expectation values:
$\om \sim {\cal O}(1)\ \mathrm{TeV},\ v \approx 246\ \mathrm{GeV}$
and $ u \sim {\cal O}(1)\ \mathrm{GeV}$. In this paper we have added
a new characteristic property, namely,  there are two types of
Yukawa couplings with different strengths: the LNC coupling $h$'s
and the LNV ones $s$'s satisfying the condition: $ s \ll h$. With
the help of these key properties, the mass spectrum of quarks is
consistent without introducing the third scalar triplet.

 With the given  set of parameters, the numerical evaluation
 shows that in this model, masses of the exotic
quarks also have different scales, namely, the $U$ exotic quark
($q_U = 2/3$) gains mass $m_U \approx 700 $ GeV, while the $D_\al$
exotic quarks ($q_{D_\al} = -1/3$) have masses in the TeV scale:
$m_{D_\al} \in 10 \div 80$ TeV.

Let us summarize our results: \ben \item {\it At the tree level}
\ben
\item  All charged leptons gain masses similar
to that in the SM.
\item One neutrino is massless and two are
degenerate in masses.
\item All exotic quarks gain masses proportional to the $\om$ - the
VEV of the first step of symmetry breaking.
\item  The quarks $u_1, d_2, d_3$ are massless
\een
\item {\it At the one-loop level}
\ben
\item  All above-mentioned quarks gain masses.
\item The light-quarks gain masses proportional to $u$ - the VEV of
lepton-number violation
\item There exist two types of Yukawa couplings: the LNC and LNV
with quite  different strengths.
\item The masses of the exotic quarks are separated too:
\be m_U \approx 700 \ \mathrm{GeV}, \ m_{D_\al} \in \ 10 \div 80 \
\mathrm{TeV} \ee \een \een
 With the {\it positive} answer, the economical version
becomes one of the very attractive models beyond the SM.

This is the time to mention some developments of the considered
model. The idea to give VEVs at the top and bottom elements of
$\chi$ triplet was given in Ref.~\cite{ponce}. Some consequences
such as the atomic parity violation, $Z-Z'$ mixing angle and $Z'$
mass were studied~\cite{ochoa}. However, in the above-mentioned
works, the $W-Y$ and $W_4, Z, Z'$ mixings were excluded. To solve
the difficulties such as the SM coupling $ZZ h$ or quark masses, the
third scalar triplet was introduced. Thus, the scalar sector did not
to be minimal anymore and the economical was though to be
unrealistic!

In the beginning of this year, there is a new step in development of
the model. In Ref~\cite{haihiggs}, the correct identification of
non-Hermitian bilepton gauge boson $X^0$ was established. The $W-Y$
mixing as well as $W_4, Z, Z'$ one were also entered into couplings
among fermions and gauge bosons. The scalar sector was studied in
Ref.~\cite{higgseconom} and all gauge-Higgs couplings were presented
and all  similar  ones in the SM  were recovered. The Higgs sector
contains eight Goldstone bosons - the needed number for massive
gauge ones of the model. Interesting to note that, the $CP$-odd part
of Goldstone associated with the neutral non-Hermitian bilepton
gauge boson $G_{X^0}$ is decouple, while its $CP$-even counterpart
has the mixing by the same way in the gauge boson sector.

In Ref.~\cite{dln}, the deviation $\de Q_{\mathrm{W}}$ of the weak
charge from its SM prediction due to the mixing of the $W$ boson
with the charged bilepton $Y$ as well as of the $Z$ boson with the
neutral $Z'$ and the real part of the non-Hermitian neutral bilepton
$X^0$ is established.

In this model, the lepton-number violating interactions exist in
both charged and neutral gauge boson sectors. However, the
lepton-number violation happens only in the neutrino and exotic
quarks sectors,  but not in the charged lepton sector. In this
model, lepton-number changing ($\De L = \pm 2$) processes exist but
{\it only} in the neutrino sector. Consequently, neutrinos get
Majorana masses at the one-loop level.

It is worth  mentioning on  the advantage of the considered model:
the new mixing angle between the charged gauge bosons $\theta$ is
connected with one of the VEVs $u$ - the parameter of lepton-number
violations. There is no new  parameter, but it contains very simple
Higgs sector, hence the significant number of free parameters is
reduced.

The model contains new kinds of interactions in the neutrino sector,
and the exotic quark  sector is rich too.  Hence
the model  deserves further studies.

\section*{Acknowledgement}
   Financial support from Swedish
International Development Cooperation Agency (SIDA) through the
Associateship Scheme of the Abdus Salam International Centre for
Theoretical Physics, Trieste, Italy is acknowledged (H. N. L.). This
work was supported in part by National Council for
Natural Sciences of Vietnam.\\[0.3cm]

\appendix
\renewcommand{\thesection}{Appendix {\Alph{section}}}
\section{  Feynman integration}
\renewcommand{\thesection}{{\Alph{section}}}
\makeatletter \setcounter{secnumdepth}{5} \setcounter{tocdepth}{5}
\renewcommand{\theequation}{\thesection.\arabic{equation}}
\@addtoreset{equation}{section} \makeatother \label{inter}

With the help of \bea &&
\int\fr{d^4p}{(2\pi)^4}\fr{1}{(p^2-a)(p^2-b)(p^2-c)}=\fr{-i}{16\pi^2}\left\{\fr{a\ln
a}{(a-b)(a-c)}\right. \crn && \left.+\fr{b\ln
b}{(b-a)(b-c)}+\fr{c\ln c}{(c-b)(c-a)}\right\},\label{eqn7031}\eea
and by differentiating two sides with respect to $a$, we have \bea
&&
\int\fr{d^4p}{(2\pi)^4}\fr{1}{(p^2-a)^2(p^2-b)(p^2-c)}=\fr{-i}{16\pi^2}\crn
&& \times\left\{\fr{\ln a +1}{(a-b)(a-c)}-\fr{a(2a -b -c)\ln a
}{(a-b)^2(a-c)^2}+\fr{b\ln b}{(b-a)^2(b-c)}+\fr{c\ln
c}{(c-a)^2(c-b)}\right\}.\label{eqn7032}\eea Hence we get \bea
I(a,b,c)&=&
\int\fr{d^4p}{(2\pi)^4}\fr{p^2}{(p^2-a)^2(p^2-b)(p^2-c)}=
\int\fr{d^4p}{(2\pi)^4}\left[\fr{1}{(p^2-a)(p^2-b)(p^2-c)}\right.\crn
&& + \left.\fr{a}{(p^2-a)^2(p^2-b)(p^2-c)}\right]\crn&=&
\fr{-i}{16\pi^2}\left\{\fr{a(2\ln a +1)}{(a-b)(a-c)} -\fr{a^2(2a -b
-c)\ln a }{(a-b)^2(a-c)^2} \right.\crn &&+\left. \fr{b^2\ln
b}{(b-a)^2(b-c)}+\fr{c^2\ln
c}{(c-a)^2(c-b)}\right\}.\label{eqn7033}\eea

In the case of $b=c$, we have also \be I(a,b,b)=-\fr{i}{16
\pi^2}\left[\fr{a+b}{(a-b)^2}-\fr{2ab}{(a-b)^3}\ln\fr{a}{b}\right].\ee

If $a,b\gg c$ we have an approximation: \be I(a,b,c)\simeq
\fr{-i}{16\pi^2}\left[\fr{a-b+b\ln\fr{b}{a}}{(a-b)^2}\right].
\label{c4}\ee

\section{One-loop corrections}

\begin{figure}[htbp]
\begin{center}
\begin{picture}(650,500)(-45,-440)

\ArrowLine(10,20)(40,20) \ArrowLine(130,20)(160,20)
\ArrowLine(40,20)(85,20) \ArrowLine(85,20)(130,20)
\DashArrowArcn(85,20)(45,90,0){2}
\DashArrowArcn(85,20)(45,180,90){2}

\Text(160,10)[]{$Q^c_{1 L}$} \Text(10,10)[]{$ u_{i R} $}
\Text(65,10)[]{$ Q_{\al b L}$} \Text(105,10)[]{$ D_{\beta R}$}
\Text(40,10)[]{$h^u_{ \al i}$}
\DashArrowLine(85,-10)(85,20){2}\Text(85,-10)[]{$\times$}
\Text(98,-10)[]{$\chi_e$} \Text(85,30)[]{$h^D_{\beta \al}$}
 \Text(130,10)[]{$s^D_{\beta} $}
\Text(133,45)[]{$ \phi_d$}
 \Text(44,50)[]{$ \phi^a$}
\Text(85,55)[]{$  \la_{4,2,3}$} \Text(85,-25)[]{(a)}
\DashArrowLine(85,65)(120,90){2} \Text(120,90)[]{$\times$}
\DashArrowLine(50,90)(85,65){2} \Text(50,90)[]{$\times$}
\Text(45,100)[]{$\chi_g, \phi_g$} \Text(120,100)[]{$\chi_h,\phi_h$}


\ArrowLine(210,20)(240,20) \ArrowLine(330,20)(360,20)
\ArrowLine(240,20)(285,20) \ArrowLine(285,20)(330,20)
\DashArrowArcn(285,20)(45,90,0){2}
\DashArrowArc(285,20)(45,90,180){2} \Text(360,10)[]{$Q^c_{1L}$}
\Text(298,-10)[]{$\chi_e$} \Text(210,10)[]{$ u_{iR} $}
\Text(260,10)[]{$ Q^b_{1L}$} \Text(305,10)[]{$ U_{R}$}
\Text(240,10)[]{$s^u_{i}$}
\DashArrowLine(285,20)(285,-10){2}\Text(285,-10)[]{$\times$}
\Text(285,30)[]{$h^U$}
 \Text(330,10)[]{$h^U $}
\Text(333,45)[]{$ \chi_d$} 
\Text(244,50)[]{$ \chi_a$} \DashArrowLine(320,90)(285,65){2}
\DashArrowLine(250,90)(285,65){2} \Text(320,90)[]{$\times$}
\Text(250,90)[]{$\times$} \Text(245,100)[]{$\chi_g$}
\Text(320,100)[]{$\chi_h$}\Text(285,-25)[]{(b)}

\Text(285,55)[]{$  \la_{1}$}


\ArrowLine(10,-120)(40,-120) \ArrowLine(130,-120)(160,-120)
\ArrowLine(40,-120)(85,-120) \ArrowLine(85,-120)(130,-120)
\DashArrowArcn(85,-120)(45,90,0){2}
\DashArrowArc(85,-120)(45,90,180){2} \Text(160,-130)[]{$Q^c_{1L}$}
\Text(10,-130)[]{$ U_{R} $}
\Text(60,-130)[]{$ Q^b_{1L}$} \Text(105,-130)[]{$ U_{R}$}
\Text(40,-130)[]{$h^U$} \DashArrowLine(85,-120)(85,-150){2}
\Text(100,-150)[]{$\chi_e$} \Text(85,-150)[]{$\times$}
\Text(85,-110)[]{$h^U$}
 \Text(130,-130)[]{$h^U $}
\Text(133,-95)[]{$ \chi_d$}
\Text(44,-90)[]{$ \chi_a$}
\Text(85,-85)[]{$  \la_{1}$} \Text(85,-165)[]{(c)}
\DashArrowLine(120,-50)(85,-75){2} \Text(120,-50)[]{$\times$}
\DashArrowLine(50,-50)(85,-75){2} \Text(50,-50)[]{$\times$}
\Text(45,-40)[]{$\chi_g$} \Text(115,-40)[]{$\chi_h$}


\ArrowLine(210,-120)(240,-120) \ArrowLine(330,-120)(360,-120)
\ArrowLine(240,-120)(285,-120) \ArrowLine(285,-120)(330,-120)
\DashArrowArcn(285,-120)(45,90,0){2}
\DashArrowArcn(285,-120)(45,180,90){2} \Text(360,-130)[]{$Q^c_{1L}$}
\Text(210,-130)[]{$ U_{R} $} \Text(260,-130)[]{$ Q_{\al b L}$}
\Text(305,-130)[]{$ D_{\beta R}$} \Text(240,-130)[]{$s^U_\al$}
\Text(285,-150)[]{$\times$} \DashArrowLine(285,-150)(285,-120){2}
\Text(285,-110)[]{$h^D_{\beta \al}$}
 \Text(330,-130)[]{$s^D_\beta $}
\Text(333,-95)[]{$ \phi_d$} 
\Text(244,-90)[]{$ \phi^a$} \DashArrowLine(250,-50)(285,-75){2}
\DashArrowLine(285,-75)(320,-50){2} \Text(320,-50)[]{$\times$}
\Text(250,-50)[]{$\times$} \Text(250,-40)[]{$\chi_g,\phi_g$}
\Text(320,-40)[]{$\chi_h,\phi_h$}\Text(285,-165)[]{(d)}
\Text(300,-150)[]{$\chi_e$} \Text(285,-85)[]{$  \la_{4,2,3}$}


\ArrowLine(10,-260)(40,-260) \ArrowLine(130,-260)(160,-260)
\ArrowLine(40,-260)(85,-260) \ArrowLine(85,-260)(130,-260)
\DashArrowArc(85,-260)(45,0,90){2}
\DashArrowArcn(85,-260)(45,180,90){2} \Text(160,-270)[]{$Q_{\ga c
L}$} \Text(10,-270)[]{$ u_{i R} $}
\Text(65,-270)[]{$ Q_{\al b L}$} \Text(105,-270)[]{$ D_{\beta R}$}
\Text(40,-270)[]{$h^u_{ \al i}$}
\DashArrowLine(85,-290)(85,-260){2}\Text(85,-290)[]{$\times$}
\Text(98,-290)[]{$\chi_e$} \Text(85,-250)[]{$h^D_{\beta \al}$}
 \Text(130,-270)[]{$h^D_{\ga \beta} $}
\Text(133,-235)[]{$ \chi^d$}
 \Text(44,-230)[]{$ \phi^a$}
\Text(85,-225)[]{$  \la_{4}$} \Text(85,-305)[]{(e)}
\DashArrowLine(85,-215)(120,-190){2} \Text(120,-190)[]{$\times$}
\DashArrowLine(85,-215)(50,-190){2} \Text(50,-190)[]{$\times$}
\Text(45,-180)[]{$\chi_g$} \Text(120,-180)[]{$\phi_h$}


\ArrowLine(210,-260)(240,-260) \ArrowLine(330,-260)(360,-260)
\ArrowLine(240,-260)(285,-260) \ArrowLine(285,-260)(330,-260)
\DashArrowArc(285,-260)(45,0,90){2}
\DashArrowArc(285,-260)(45,90,180){2} \Text(360,-270)[]{$Q_{\ga c
L}$} \Text(298,-290)[]{$\chi_e$} \Text(210,-270)[]{$ u_{iR} $}
\Text(260,-270)[]{$ Q^b_{1L}$} \Text(305,-270)[]{$ U_{R}$}
\Text(240,-270)[]{$s^u_{i}$}
\DashArrowLine(285,-260)(285,-290){2}\Text(285,-290)[]{$\times$}
\Text(285,-250)[]{$h^U$}
 \Text(330,-270)[]{$s^U_\ga $}
\Text(333,-235)[]{$ \phi^d$} 
\Text(244,-230)[]{$ \chi_a$} \DashArrowLine(320,-190)(285,-215){2}
\DashArrowLine(285,-215)(250,-190){2} \Text(320,-190)[]{$\times$}
\Text(250,-190)[]{$\times$} \Text(245,-180)[]{$\phi_g$}
\Text(320,-180)[]{$\chi_h$}\Text(285,-305)[]{(f)}

\Text(285,-225)[]{$  \la_{3}$}


\ArrowLine(10,-400)(40,-400) \ArrowLine(130,-400)(160,-400)
\ArrowLine(40,-400)(85,-400) \ArrowLine(85,-400)(130,-400)
\DashArrowArc(85,-400)(45,0,90){2}
\DashArrowArc(85,-400)(45,90,180){2} \Text(160,-410)[]{$Q_{\ga c
L}$} \Text(10,-410)[]{$ U_{R} $}
\Text(60,-410)[]{$ Q^b_{1L}$} \Text(105,-410)[]{$ U_{R}$}
\Text(40,-410)[]{$h^U$} \DashArrowLine(85,-400)(85,-430){2}
\Text(100,-430)[]{$\chi_e$} \Text(85,-430)[]{$\times$}
\Text(85,-390)[]{$h^U$}
 \Text(130,-410)[]{$s^U_\ga $}
\Text(133,-375)[]{$ \phi^d$} \Text(44,-370)[]{$ \chi_a$}
\Text(85,-365)[]{$  \la_{3}$} \Text(85,-445)[]{(g)}
\DashArrowLine(120,-330)(85,-355){2} \Text(120,-330)[]{$\times$}
\DashArrowLine(85,-355)(50,-330){2} \Text(50,-330)[]{$\times$}
\Text(45,-320)[]{$\phi_g$} \Text(115,-320)[]{$\chi_h$}


\ArrowLine(210,-400)(240,-400) \ArrowLine(330,-400)(360,-400)
\ArrowLine(240,-400)(285,-400) \ArrowLine(285,-400)(330,-400)
\DashArrowArc(285,-400)(45,0,90){2}
\DashArrowArcn(285,-400)(45,180,90){2} \Text(360,-410)[]{$Q_{\ga c
L}$} \Text(210,-410)[]{$ U_{R} $} \Text(260,-410)[]{$ Q_{\al b L}$}
\Text(305,-410)[]{$ D_{\beta R}$} \Text(240,-410)[]{$s^U_\al$}
\Text(285,-430)[]{$\times$} \DashArrowLine(285,-430)(285,-400){2}
\Text(285,-390)[]{$h^D_{\beta \al}$}
 \Text(330,-410)[]{$h^D_{\ga \beta} $}
\Text(333,-375)[]{$ \chi^d$} 
\Text(244,-370)[]{$ \phi^a$} \DashArrowLine(285,-355)(250,-330){2}
\DashArrowLine(285,-355)(320,-330){2} \Text(320,-330)[]{$\times$}
\Text(250,-330)[]{$\times$} \Text(250,-320)[]{$\chi_g$}
\Text(320,-320)[]{$\phi_h$}\Text(285,-445)[]{(h)}
\Text(300,-430)[]{$\chi_e$} \Text(285,-365)[]{$  \la_{4}$}
\Text(100,-465)[]{$+\ 16\ \mathrm{ graphs\ with\ smaller\
contributions}$}

\end{picture}
\end{center}
\vs
 \caption[]{\label{figh4} One-loop contributions to the up-quark
mass matrix (\ref{upqmasstu}). }
\end{figure}
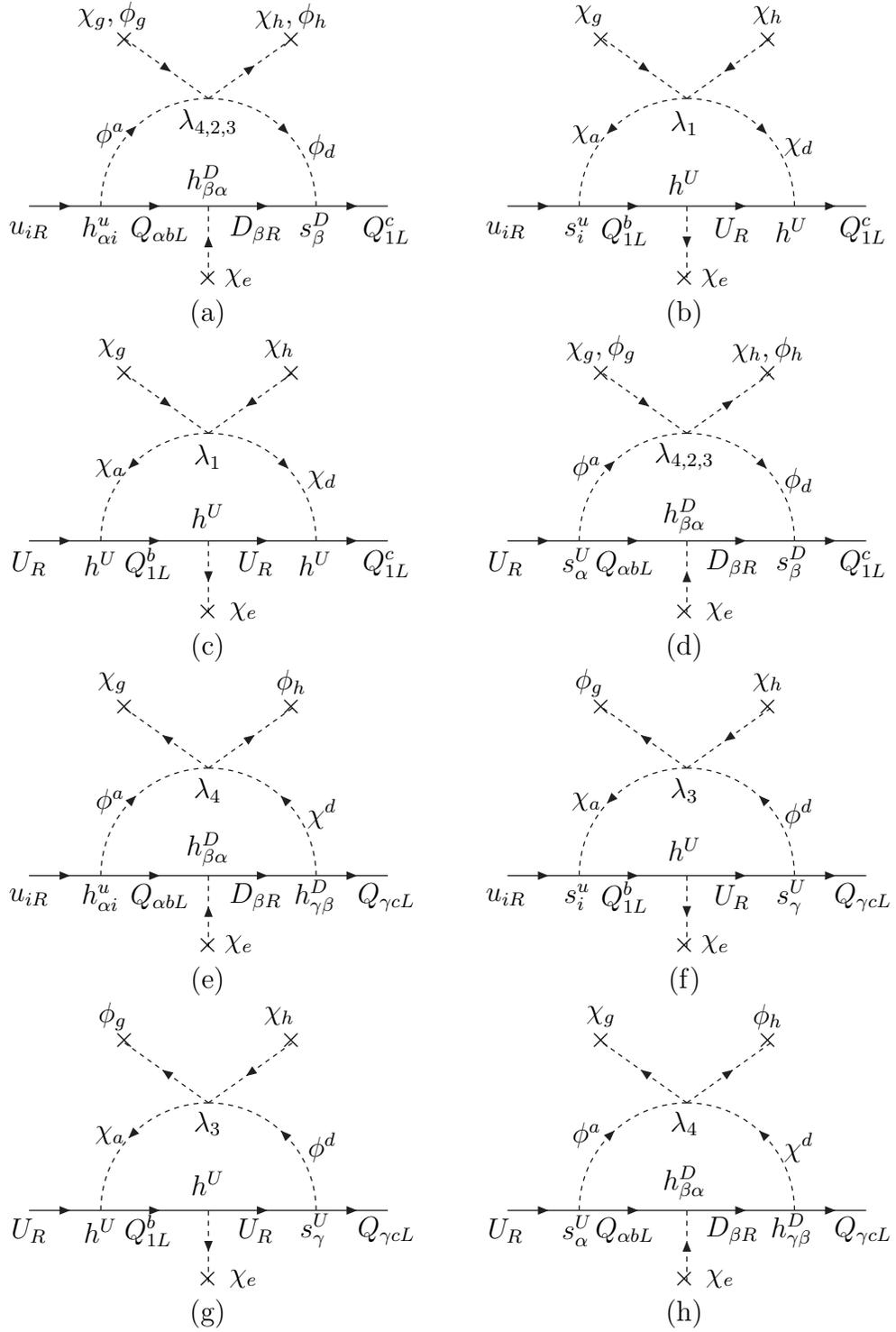

\begin{figure}[htbp]
\begin{center}
\begin{picture}(650,500)(-45,-440)

\ArrowLine(10,20)(40,20) \ArrowLine(130,20)(160,20)
\ArrowLine(40,20)(85,20) \ArrowLine(85,20)(130,20)
\DashArrowArc(85,20)(45,0,90){2}
\DashArrowArcn(85,20)(45,180,90){2}

\Text(160,10)[]{$Q_{\ga c L}$} \Text(10,10)[]{$ d_{i R} $}
\Text(65,10)[]{$ Q_{\al b L}$} \Text(105,10)[]{$ D_{\beta R}$}
\Text(40,10)[]{$s^d_{ \al i}$}
\DashArrowLine(85,-10)(85,20){2}\Text(85,-10)[]{$\times$}
\Text(98,-10)[]{$\chi_e$} \Text(85,30)[]{$h^D_{\beta \al}$}
 \Text(130,10)[]{$h^D_{\ga \beta} $}
\Text(133,45)[]{$ \chi^d$}
 \Text(44,50)[]{$ \chi^a$}
\Text(85,55)[]{$  \la_{1}$} \Text(85,-25)[]{(a)}
\DashArrowLine(85,65)(120,90){2} \Text(120,90)[]{$\times$}
\DashArrowLine(85,65)(50,90){2} \Text(50,90)[]{$\times$}
\Text(45,100)[]{$\chi_g$} \Text(120,100)[]{$\chi_h$}


\ArrowLine(210,20)(240,20) \ArrowLine(330,20)(360,20)
\ArrowLine(240,20)(285,20) \ArrowLine(285,20)(330,20)
\DashArrowArc(285,20)(45,0,90){2}
\DashArrowArc(285,20)(45,90,180){2} \Text(360,10)[]{$Q_{\ga cL}$}
\Text(298,-10)[]{$\chi_e$} \Text(210,10)[]{$ d_{iR} $}
\Text(260,10)[]{$ Q^b_{1L}$} \Text(305,10)[]{$ U_{R}$}
\Text(240,10)[]{$h^d_{i}$}
\DashArrowLine(285,20)(285,-10){2}\Text(285,-10)[]{$\times$}
\Text(285,30)[]{$h^U$}
 \Text(330,10)[]{$s^U_\ga $}
\Text(333,45)[]{$ \phi^d$} 
\Text(244,50)[]{$ \phi_a$} \DashArrowLine(320,90)(285,65){2}
\DashArrowLine(285,65)(250,90){2} \Text(320,90)[]{$\times$}
\Text(250,90)[]{$\times$} \Text(245,100)[]{$\chi_g,\phi$}
\Text(320,100)[]{$\chi_h,\phi$}\Text(285,-25)[]{(b)}

\Text(285,55)[]{$  \la_{4,2,3}$}

\ArrowLine(10,-120)(40,-120) \ArrowLine(130,-120)(160,-120)
\ArrowLine(40,-120)(85,-120) \ArrowLine(85,-120)(130,-120)
\DashArrowArc(85,-120)(45,0,90){2}
\DashArrowArc(85,-120)(45,90,180){2} \Text(160,-130)[]{$Q_{\ga c
L}$} \Text(10,-130)[]{$ D_{\de R} $}
\Text(60,-130)[]{$ Q^b_{1L}$} \Text(105,-130)[]{$ U_{R}$}
\Text(40,-130)[]{$s^D_\de$} \DashArrowLine(85,-120)(85,-150){2}
\Text(100,-150)[]{$\chi_e$} \Text(85,-150)[]{$\times$}
\Text(85,-110)[]{$h^U$}
 \Text(130,-130)[]{$s^U_\ga $}
\Text(133,-95)[]{$ \phi^d$} \Text(44,-90)[]{$ \phi_a$}
\Text(85,-85)[]{$  \la_{4,2,3}$} \Text(85,-165)[]{(c)}
\DashArrowLine(120,-50)(85,-75){2} \Text(120,-50)[]{$\times$}
\DashArrowLine(85,-75)(50,-50){2} \Text(50,-50)[]{$\times$}
\Text(45,-40)[]{$\chi_g,\phi_g$} \Text(115,-40)[]{$\chi_h,\phi_h$}


\ArrowLine(210,-120)(240,-120) \ArrowLine(330,-120)(360,-120)
\ArrowLine(240,-120)(285,-120) \ArrowLine(285,-120)(330,-120)
\DashArrowArc(285,-120)(45,0,90){2}
\DashArrowArcn(285,-120)(45,180,90){2} \Text(360,-130)[]{$Q_{\ga c
L}$} \Text(210,-130)[]{$ D_{\de R} $} \Text(260,-130)[]{$ Q_{\al b
L}$} \Text(305,-130)[]{$ D_{\beta R}$} \Text(240,-130)[]{$h^D_{\al
\de}$} \Text(285,-150)[]{$\times$}
\DashArrowLine(285,-150)(285,-120){2} \Text(285,-110)[]{$h^D_{\beta
\al}$}
 \Text(330,-130)[]{$h^D_{\ga \beta} $}
\Text(333,-95)[]{$ \chi^d$} 
\Text(244,-90)[]{$ \chi^a$} \DashArrowLine(285,-75)(250,-50){2}
\DashArrowLine(285,-75)(320,-50){2} \Text(320,-50)[]{$\times$}
\Text(250,-50)[]{$\times$} \Text(250,-40)[]{$\chi_g$}
\Text(320,-40)[]{$\chi_h$}\Text(285,-165)[]{(d)}
\Text(300,-150)[]{$\chi_e$} \Text(285,-85)[]{$  \la_{1}$}


\ArrowLine(10,-260)(40,-260) \ArrowLine(130,-260)(160,-260)
\ArrowLine(40,-260)(85,-260) \ArrowLine(85,-260)(130,-260)
\DashArrowArcn(85,-260)(45,90,0){2}
\DashArrowArcn(85,-260)(45,180,90){2} \Text(160,-270)[]{$Q^c_{1L}$}
\Text(10,-270)[]{$ d_{i R} $}
\Text(65,-270)[]{$ Q_{\al b L}$} \Text(105,-270)[]{$ D_{\beta R}$}
\Text(40,-270)[]{$s^d_{ \al i}$}
\DashArrowLine(85,-290)(85,-260){2}\Text(85,-290)[]{$\times$}
\Text(98,-290)[]{$\chi_e$} \Text(85,-250)[]{$h^D_{\beta \al}$}
 \Text(130,-270)[]{$s^D_{\beta} $}
\Text(133,-235)[]{$ \phi_d$}
 \Text(44,-230)[]{$ \chi^a$}
\Text(85,-225)[]{$  \la_{3}$} \Text(85,-305)[]{(e)}
\DashArrowLine(85,-215)(120,-190){2} \Text(120,-190)[]{$\times$}
\DashArrowLine(50,-190)(85,-215){2} \Text(50,-190)[]{$\times$}
\Text(45,-180)[]{$\phi_g$} \Text(120,-180)[]{$\chi_h$}


\ArrowLine(210,-260)(240,-260) \ArrowLine(330,-260)(360,-260)
\ArrowLine(240,-260)(285,-260) \ArrowLine(285,-260)(330,-260)
\DashArrowArcn(285,-260)(45,90,0){2}
\DashArrowArc(285,-260)(45,90,180){2} \Text(360,-270)[]{$Q^c_{1 L}$}
\Text(298,-290)[]{$\chi_e$} \Text(210,-270)[]{$ d_{iR} $}
\Text(260,-270)[]{$ Q^b_{1L}$} \Text(305,-270)[]{$ U_{R}$}
\Text(240,-270)[]{$h^d_{i}$}
\DashArrowLine(285,-260)(285,-290){2}\Text(285,-290)[]{$\times$}
\Text(285,-250)[]{$h^U$}
 \Text(330,-270)[]{$h^U$}
\Text(333,-235)[]{$ \chi_d$} 
\Text(244,-230)[]{$ \phi_a$} \DashArrowLine(320,-190)(285,-215){2}
\DashArrowLine(250,-190)(285,-215){2} \Text(320,-190)[]{$\times$}
\Text(250,-190)[]{$\times$} \Text(245,-180)[]{$\chi_g$}
\Text(320,-180)[]{$\phi_h$}\Text(285,-305)[]{(f)}

\Text(285,-225)[]{$  \la_{4}$}


\ArrowLine(10,-400)(40,-400) \ArrowLine(130,-400)(160,-400)
\ArrowLine(40,-400)(85,-400) \ArrowLine(85,-400)(130,-400)
\DashArrowArcn(85,-400)(45,90,0){2}
\DashArrowArc(85,-400)(45,90,180){2} \Text(160,-410)[]{$Q^c_{1 L}$}
\Text(10,-410)[]{$ D_{\de R} $}
\Text(60,-410)[]{$ Q^b_{1L}$} \Text(105,-410)[]{$ U_{R}$}
\Text(40,-410)[]{$s^D_\de$} \DashArrowLine(85,-400)(85,-430){2}
\Text(100,-430)[]{$\chi_e$} \Text(85,-430)[]{$\times$}
\Text(85,-390)[]{$h^U$}
 \Text(130,-410)[]{$h^U $}
\Text(133,-375)[]{$ \chi_d$} \Text(44,-370)[]{$ \phi_a$}
\Text(85,-365)[]{$  \la_{4}$} \Text(85,-445)[]{(g)}
\DashArrowLine(120,-330)(85,-355){2} \Text(120,-330)[]{$\times$}
\DashArrowLine(50,-330)(85,-355){2} \Text(50,-330)[]{$\times$}
\Text(45,-320)[]{$\chi_g$} \Text(115,-320)[]{$\phi_h$}


\ArrowLine(210,-400)(240,-400) \ArrowLine(330,-400)(360,-400)
\ArrowLine(240,-400)(285,-400) \ArrowLine(285,-400)(330,-400)
\DashArrowArcn(285,-400)(45,90,0){2}
\DashArrowArcn(285,-400)(45,180,90){2} \Text(360,-410)[]{$Q^c_{1
L}$} \Text(210,-410)[]{$ D_{\de R} $} \Text(260,-410)[]{$ Q_{\al b
L}$} \Text(305,-410)[]{$ D_{\beta R}$}
\Text(240,-410)[]{$h^D_{\al\de}$} \Text(285,-430)[]{$\times$}
\DashArrowLine(285,-430)(285,-400){2} \Text(285,-390)[]{$h^D_{\beta
\al}$}
 \Text(330,-410)[]{$s^D_{\beta} $}
\Text(333,-375)[]{$ \phi_d$} 
\Text(244,-370)[]{$ \chi^a$} \DashArrowLine(250,-330)(285,-355){2}
\DashArrowLine(285,-355)(320,-330){2} \Text(320,-330)[]{$\times$}
\Text(250,-330)[]{$\times$} \Text(250,-320)[]{$\phi_g$}
\Text(320,-320)[]{$\chi_h$}\Text(285,-445)[]{(h)}
\Text(300,-430)[]{$\chi_e$} \Text(285,-365)[]{$  \la_{3}$}
\Text(100,-465)[]{$+\ 16\ \mathrm{ graphs\ with\ smaller\
contributions}$}

\end{picture}
\end{center}
\vs

 \caption[]{\label{figh5} One-loop contributions to the
down-quark mass matrix (\ref{upqmasstdow2}). }
\end{figure}
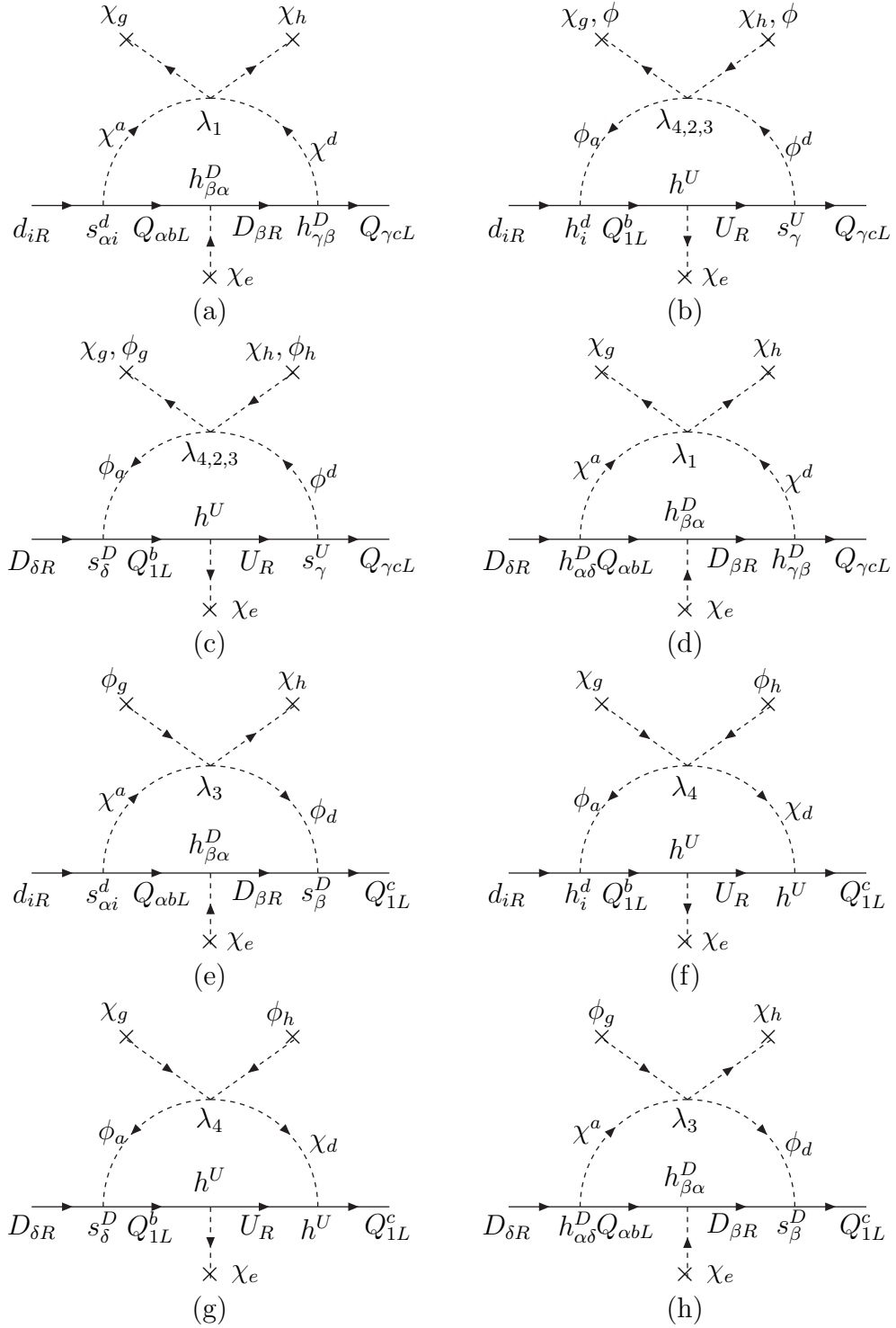

\vs

\end{document}